\documentclass[3p,times,twocolumn]{elsarticle}

\usepackage{ecrc}

\volume{00}

\firstpage{1}

\journalname{Nuclear and Particle Physics Proceedings}

\runauth{}

\jid{nppp}

\jnltitlelogo{Nuclear and Particle Physics Proceedings}

\usepackage{amssymb}
\usepackage{amsmath}
\usepackage{color}
\usepackage{soul}

\usepackage[figuresright]{rotating}

\begin{document}

\begin{frontmatter}



\dochead{}

\title{PHENIX results on direct photon-hadron correlations}


\author{Huijun Ge for the PHENIX Collaboration}

\address{}

\begin{abstract}
\setstcolor{red}
Direct photon-hadron correlations are a golden channel to study parton in-medium energy loss in QGP. The modification of the effective fragmentation function for the away-side jet can be measured by comparing integrated away-side yields of direct photon-hadron pairs in heavy ion collisions to those in p+p. We measured per-trigger-yield of associated hadrons in Au+Au collisions and observed that there is a suppression compared to p+p for higher momentum fraction ($z_T$) hadrons. This can be explained by the opacity of the hot medium: energetic partons from the initial hard scattering lose energy while traversing it. A yield enhancement on the other hand has been found at low $z_T$ (high $\xi$). Medium response is likely to be responsible for the enhanced production of these lower momentum particles. 

The same measurement is done in d+Au collisions and the result suggests that no modification to the fragmentation function is observed, given the current uncertainties.

\end{abstract}


\end{frontmatter}
%%\linenumbers

% \linenumbers

\setstcolor{red}

\section{Introduction}

In heavy-ion collisions, there are several advantages for using direct photon-hadron correlations to study parton in-medium energy loss in quark gluon plasma. First of all, since photons do not interact strongly with medium, there is no trigger surface bias compared to dihadron correlations or reconstructed jet measurement. Second of all, high momentum direct photons are produced back-to-back with partons, at leading order, from initial hard scatterings. The measured photon momentum will approximately balance that of the opposing parton before any medium modification. In that sense, using high-momentum direct photons as triggers is a most direct measure of the initial parton energy. This measurement is also complimentary to other jet measurement as it probes different path length dependence and could potentially distinguish relative contributions from quark and gluon jets. Measuring direct photon-hadron in Au+Au and comparing to the p+p baseline directly quantifies the modification of the recoil jet fragmentation function. This helps constrain models explaining soft particle production and jet-medium interactions.

The previous PHENIX measurement \cite{ppg113}  showed that there is an enhancement in soft particle production at large azimuthal angle difference $\Delta\phi$ respect to the away-side jet axis. Due to statistical limitations, it was not possible to investigate how the fragmentation function depends on the parton energy or medium scale. The motivation for this new Au+Au analysis is to add more statistics and do a differential measurement to study any possible trigger $p_T$ dependence.

The physics question we would like to address from this measurement is whether the lost energy is being fully redistributed into enhancing the production of lower momentum particles. Alternatively, some of the energy loss could be deposited into the medium and thermalized with the medium. In the latter picture, the enhancement would come mainly from jet-induced medium response. To answer this question it is crucial to observe whether $I_{AA}$ ($I_{AA} = Y_{AA}/Y_{pp}$) crossing unity is at a fixed $\xi$ or at a fixed associated particle $p_T$. The CMS jet fragmentation function measurement \cite{CMS} has shown that an enhancement to suppression starts at the region where associated particle $p_T$ is about 3-5 GeV/c, for jets $p_T$ above 100 GeV/c. 

Parton energy loss in the medium can be observed as a modification to the jet fragmentation function. With the momentum of the trigger photon as a proxy for the away-side jet momentum, the effective fragmentation function can be measured and expressed in terms of $z_T$ ( $z_T\approx p_T^h/p_T^{\gamma}$ ). To focus more on the low $z_T$ region (where the enhancement is), we can express the fragmentation function in terms of variable $\xi$ ( $\xi=ln(z_T)$ ). Measuring direct photon-hadrons in d+Au collisions probes potential cold nuclear matter effect and serves as a test for the initial state energy loss hypothesis. 
%%\label{}

\section{Experimental Setup}
The analyses are done using data collected mainly from the central arm detectors of PHENIX. The acceptance covers rapidity range of $|\eta | <$ 0.35 and azimuthal angle coverage is $\pi$. We use Electromagnetic Calorimeter (EMCal) to measure photons and $\pi^0$. Drift chambers (DC) and pad chambers (PC) reconstruct charged tracks. Beam Beam Counters (BBC) are used to determine collision centrality and vertex position. 

\section{Method}
Two particle correlations are constructed as a function of the azimuthal angle $\Delta\phi$ between triggers and associate partners. We assume pairs arise from jet correlations superimposed on a combinatorial background yield from the underlying event. In p+p and d+Au collisions where the event multiplicity is low, we assume this background is flat across $\Delta\phi$ and subtract it using the zero-yield-at-minimum (ZYAM) procedure \cite{ZYAM}. In Au+Au collisions, the background has an azimuthal asymmetry quantified in the flow parameters $v_n$, which is included in the background subtraction step, as illustrated in Eqn. 1. Note that higher order effects are included as an extra systematic uncertainty, only $v_2$ effect is included in the final results presented later.

We report jet pairs as per-trigger-yields (PTY) of hadrons. Detector acceptance is corrected using event mixing. For Au+Au collisions, the background level $b_0$ is determined using absolute normalization \cite{ZYAM}. Assuming the background is combinatorial in nature, $b_0$ can be determined using the trigger and associated partner production rates. Normalizing the yields after background subtraction by the number of triggers $N_t$ and then correct for hadron efficiency $\epsilon^a$, one could get the per-trigger-yield of associate hadrons, as described in Eq. 1. 

\begin{equation}
\begin{split}
\frac{1}{N_t}\frac{dN^{pair}}{d\Delta\phi} & = \frac{1}{N_t}\frac{N^{pair}}{\epsilon^{a}\int\Delta\phi}\Bigg\{\frac{dN^{pair}_{real}/d\Delta\phi}{dN^{pair}_{mix}/d\Delta\phi}\\
& - b_0\big[1+2\big\langle v_2^tv_2^a \big\rangle cos\big(2\Delta\phi\big) \big] \Bigg\}
\end{split}
\end{equation}

In order to extract per-trigger yields of hadrons associated with direct photons, the background from decay photon-hadron correlations needs to be subtracted. This is achieved  via a statistical subtraction procedure illustrated by Eqn. 2, in which $R_{\gamma}$ is the ratio of number of inclusive photons to number of decay photons, measured independently \cite{ppg139}. Decay photon background is estimated using a Monte Carlo simulation pair-by-pair mapping procedure. It calculates the probability of getting decay pairs within certain $p_T$ range from the measured $\pi^0$-h pair yields.

\begin{equation}
Y_{direct} = \frac{R_{\gamma}Y_{inclusive} - Y_{decay}}{R_{\gamma}-1}
\end{equation}

\section{Results}
Fig. 1 shows angular distributions of direct photon-hadron pairs for the 0-40$\%$ most central Au+Au data, compared to p+p. These are presented in $\xi$ bins. The near side yield is consistent with 0, indicating the background has been properly subtracted. The p+p points at the region of $\Delta\phi < 1$ rad are not shown, as a photon isolation cut is made on the near-side. On the away-side, there is an enhancement in Au+Au in high $\xi$ bins. As $\xi$ gets smaller, the Au+Au yield is suppressed.

\begin{figure}[h]
\label{fig:cdir_aa}
\centering\includegraphics[width=1.0\linewidth]{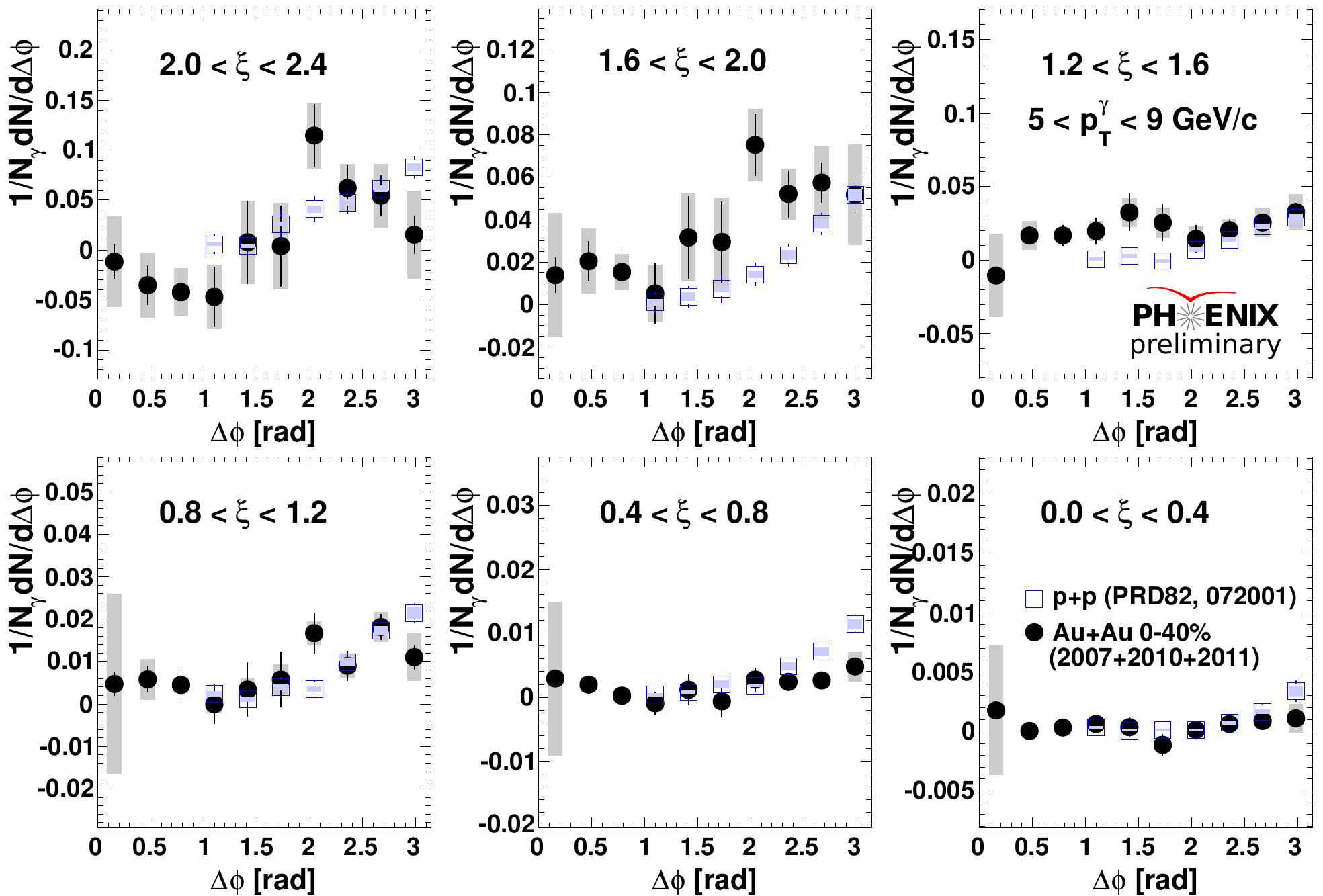}
\caption {$\Delta\phi$ distributions of direct photon-hadron pairs measured in Au+Au (black solid dots), compared with p+p baseline (blue open squares), for trigger $p_T$ range within 5-9 GeV/c, correlated with hadrons momentum corresponding to $\xi$ range of 0-2.4.}
\end{figure}

Fig. 2 shows the measured direct photon-hadron pair angular distributions in d+Au compared to p+p. Unlike in Au+Au, the d+Au correlation is consistent with the p+p baseline, given current uncertainties, indicating that any possible cold nuclear matter effect is minimal.

\begin{figure}[h]
\label{fig:cdir_da}
\centering\includegraphics[width=1.0\linewidth]{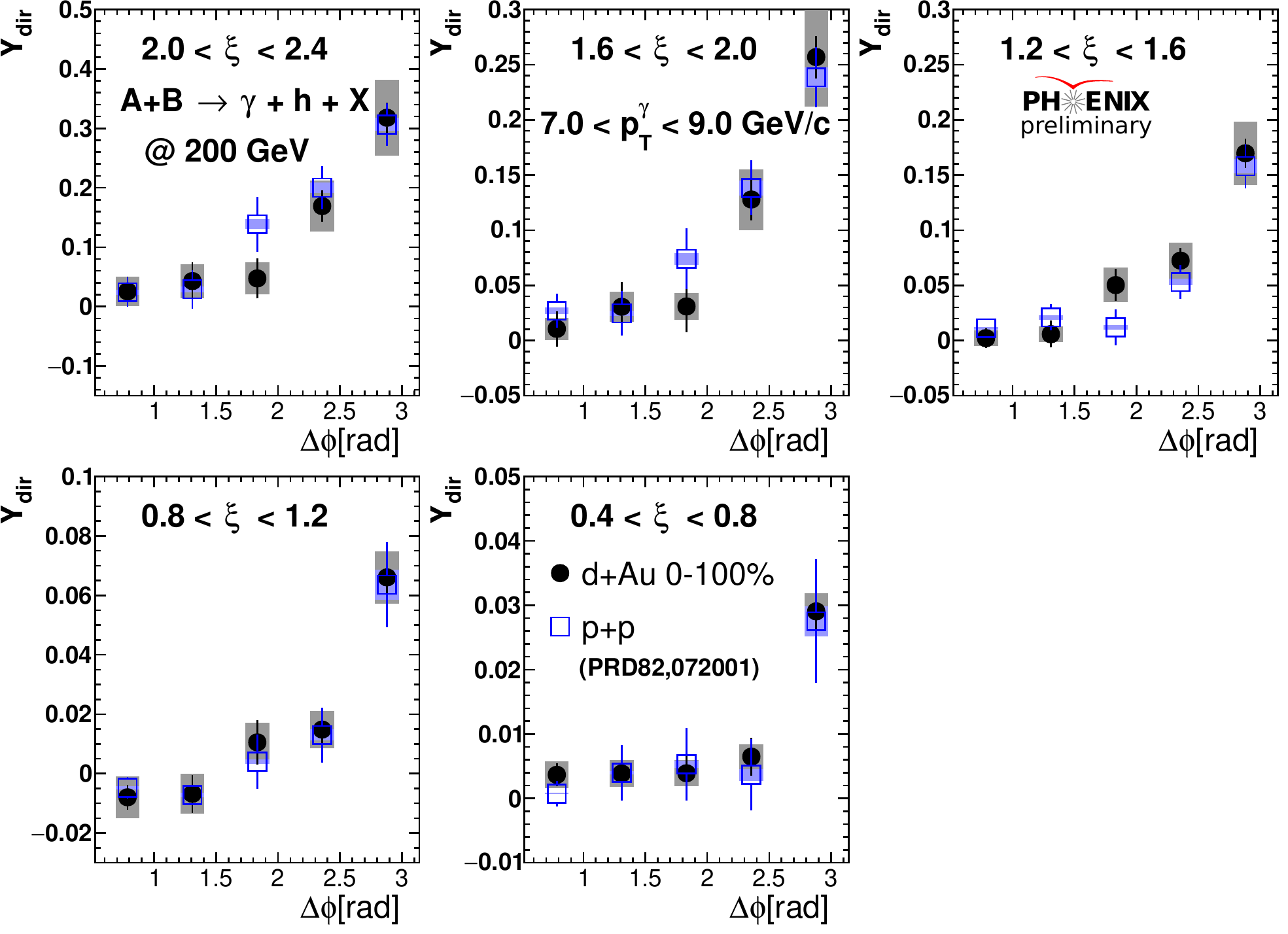}
\caption {$\Delta\phi$ distributions of direct photon-hadron pairs measured in d+Au (black solid dots), compared with p+p baseline (blue open squares), for trigger $p_T$ range within 5-9 GeV/c, correlated with hadrons momentum corresponding to $\xi$ range of 0-1.6.}
\end{figure}

We calculate the integrated away-side per-trigger yield, to get the effective fragmentation function, as shown in Fig. 3. The top panel shows yields in Au+Au, d+Au and p+p. Note that the p+p and d+Au points have been shifted to the left. Taking a ratio of the $\xi$ distribution in Au+Au to that in p+p results in $I_{AA}$, shown in the bottom panel of the figure. In the absence of modification, $I_{AA}$ would equal to 1. The data instead shows suppression at low $\xi$ and enhancement at higher $\xi$. $I_{dA}$ is also calculated and shown in the same panel ( $I_{dA} = Y_{dA}/Y_{pp}$ ). It is consistent with 1, indicating no significant medium modification in d+Au collisions. 

\begin{figure}[h]
\label{fig:yields}
\centering\includegraphics[width=1.0\linewidth]{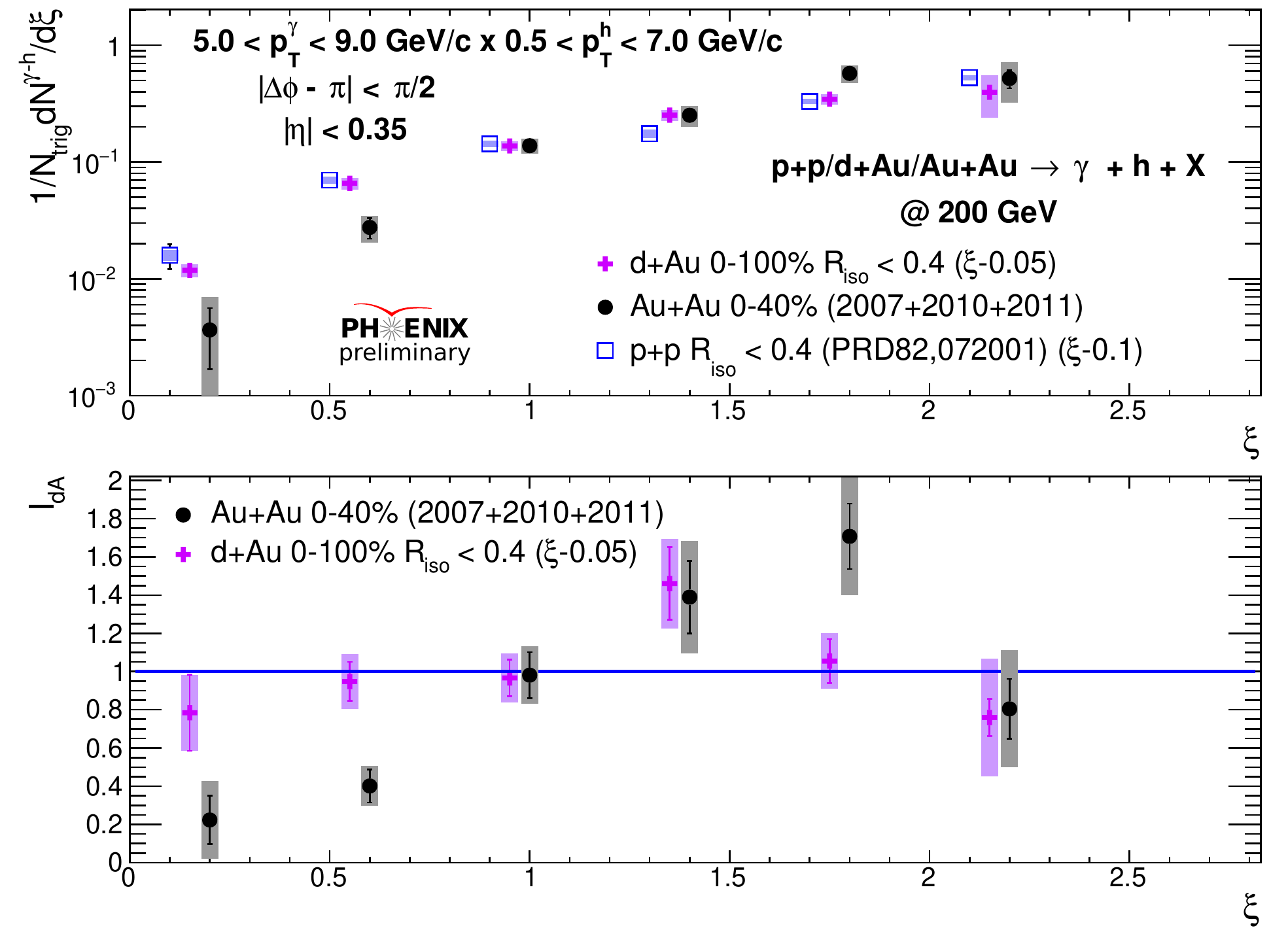}
\caption {Per-trigger-yields vs $\xi$ for Au+Au (black dots), d+Au (purple crosses) and p+p (blue open squares).}
\end{figure}

To better understand jet-medium interactions that can contribute to the modification observed in Au+Au collisions, finer trigger momentum ranges are needed. The increased statistics now allow for such differential measurement as a function of trigger $p_T$ (the jet energy). Fig. 4 shows $I_{AA}$ as a function of $\xi$ for trigger photon $p_T$ within 5-7, 7-9 and 9-12 GeV/c. While the associated hadron yields are smaller than those in p+p at low $\xi$, the appearance of extra particles at higher $\xi$ is observed for triggers with $p_T$ range of 5-7 GeV/c. A qualitatively similar rising behavior of $I_{AA}$ is also visible  for the 7-9 GeV/c $p_T$ bin.

\begin{figure}[h]
\label{fig:iaa}
\centering\includegraphics[width=1.0\linewidth]{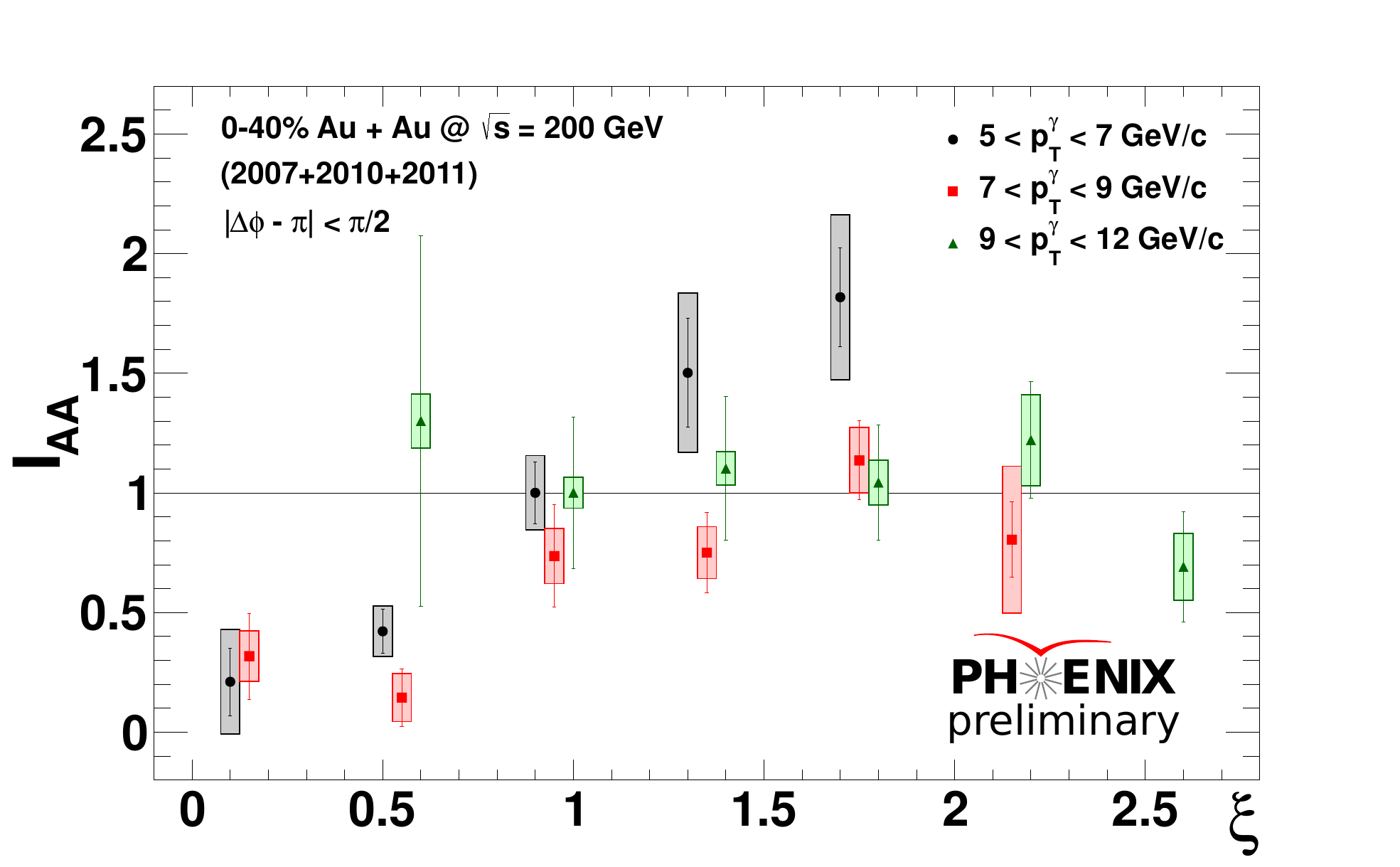}
\caption {$I_{AA}$ as a function of $\xi$ for trigger photon $p_T$ within 5-7, 7-9 and 9-12 GeV/c.}
\end{figure}

Fig.5 shows the same $I_{AA}$ as a function of $\xi$ in separate panels, compared to several theory calculations. The solid lines are CoLBT$\_$Hydro model \cite{LBT} calculated specifically for the same kinematic ranges as the data. This model is a kinetic description of parton propagation plus a hydrodynamic description of the medium evolution. It keeps track of leading partons as well as the thermal recoil partons and their further propagation. The agreement with data is reasonably good. The model clearly shows that as the trigger photon $p_T$ increases, the transition starts at a larger $\xi$ value. According to this calculation, the enhancement at large $\xi$ is from jet-induced medium excitations, therefore should have a characteristic $p_T$. Also included in Fig. 5 is a BW-MLLA calculation \cite{BW-MLLA}, shown as the dashed line in the 7-9 GeV/c trigger $p_T$ panel. This curve is calculated for jets with energy of 7 GeV. 

\begin{figure}[h]
\label{fig:theory}
\centering\includegraphics[width=1.0\linewidth]{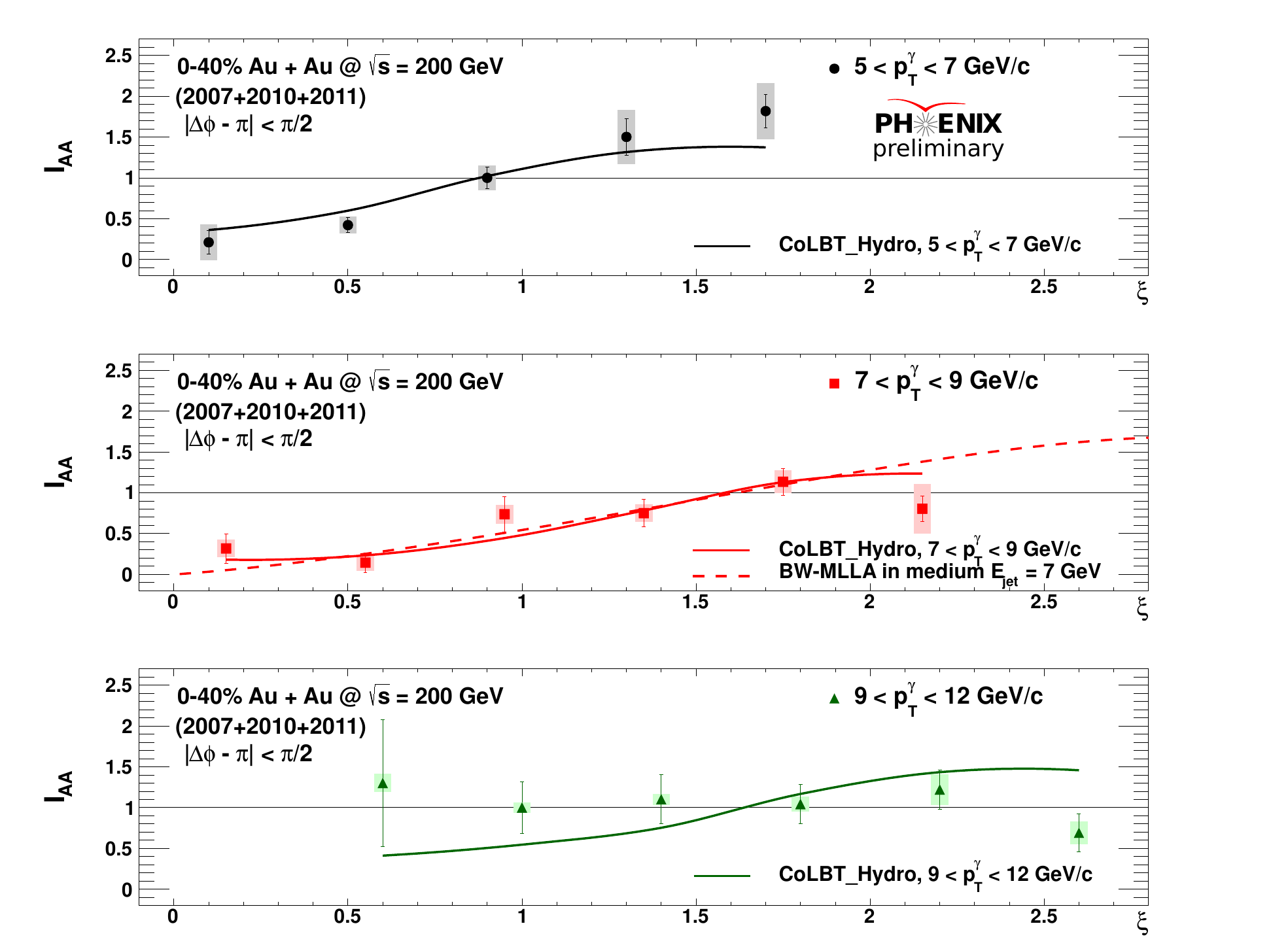}
\caption {$I_{AA}$ comparisons to model calculations.}
\end{figure}

As a further study to investigate where the energy goes, one can take a look at the $I_{AA}$ dependence on the particle distributions around the away-side jet axis. Two narrower away-side integration ranges are selected and the integrated yields are calculated. The resulting $I_{AA}$ are shown in Fig. 6. The enhancement compared to p+p is most dominant in the 5-7 GeV/c trigger $p_T$ range and for the full $\pi/2$ integration range. This can be potentially explained by the fact that lower momentum partons go through more multiple scatterings as they traverse the medium, therefore losing a larger fraction of energy to the medium. The enhancement comes from the jet-included medium response. More directly one can actually see where the particles went by looking at the angular distributions in Fig. 1: the particle yields are indeed enhanced at larger angle with respect to the away-side jet axis.

\begin{figure}[h]
\label{fig:ranges}
\centering\includegraphics[width=1.0\linewidth]{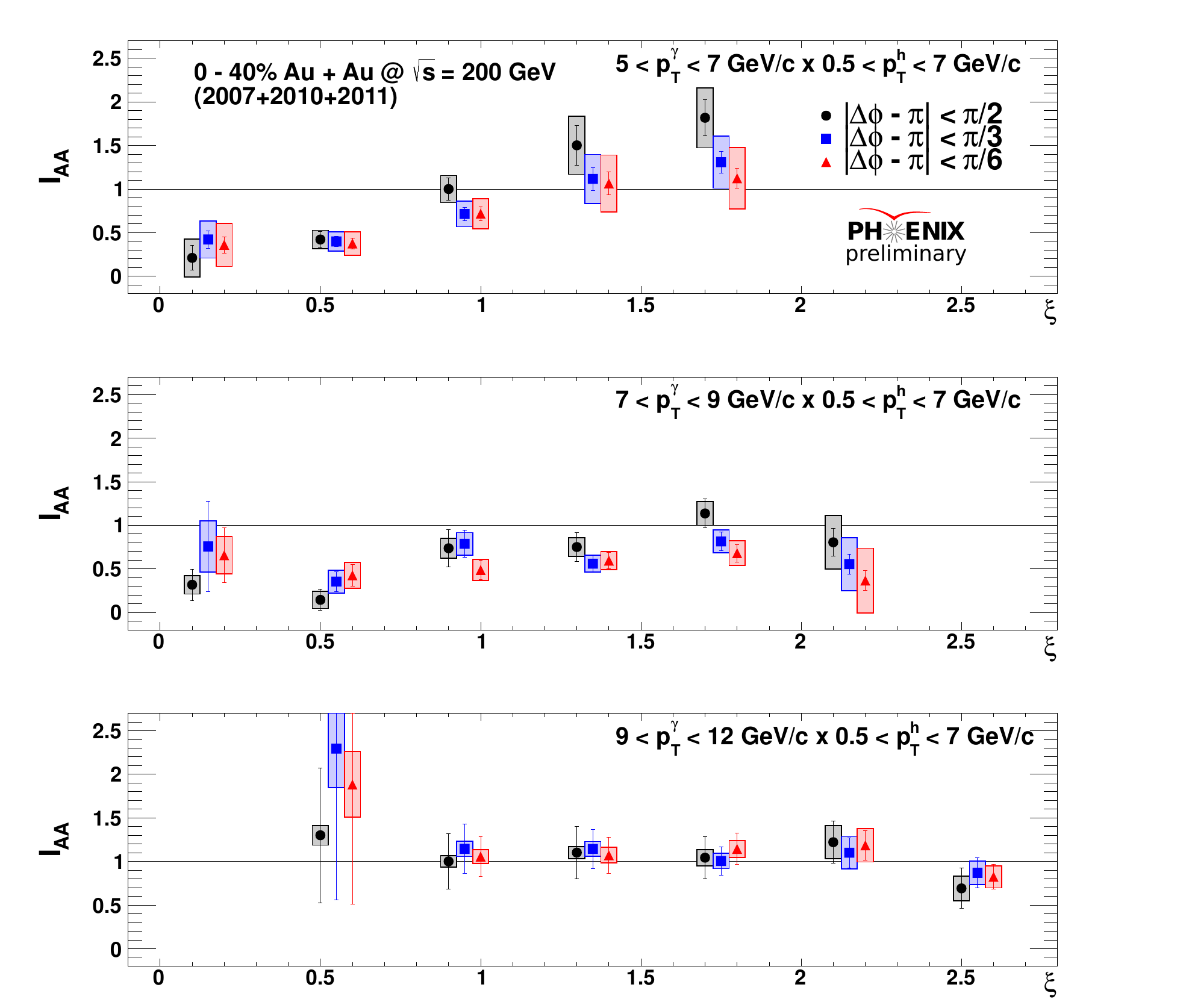}
\caption {$I_{AA}$ as a function of $\xi$ for trigger photon $p_T$ within 5-7, 7-9 and 9-12 GeV/c, for three different away-side integration ranges.}
\end{figure}

To quantify the effect of the relative enhancement at high $\xi$, compared low $\xi$, an $I_{AA}$ ratio is taken and plot as a function of the trigger photon $p_T$, shown in Fig. 7. The enhancement is most dominant for softer jets and for full away-side integration range, implying that jets with lower energy are more broadened than harder jets. This is consistent with the observation of minimal jet shape modification for very high $p_T$ jets measured at LHC.

\begin{figure}[h]
\label{fig:ranges}
\centering\includegraphics[width=1.0\linewidth]{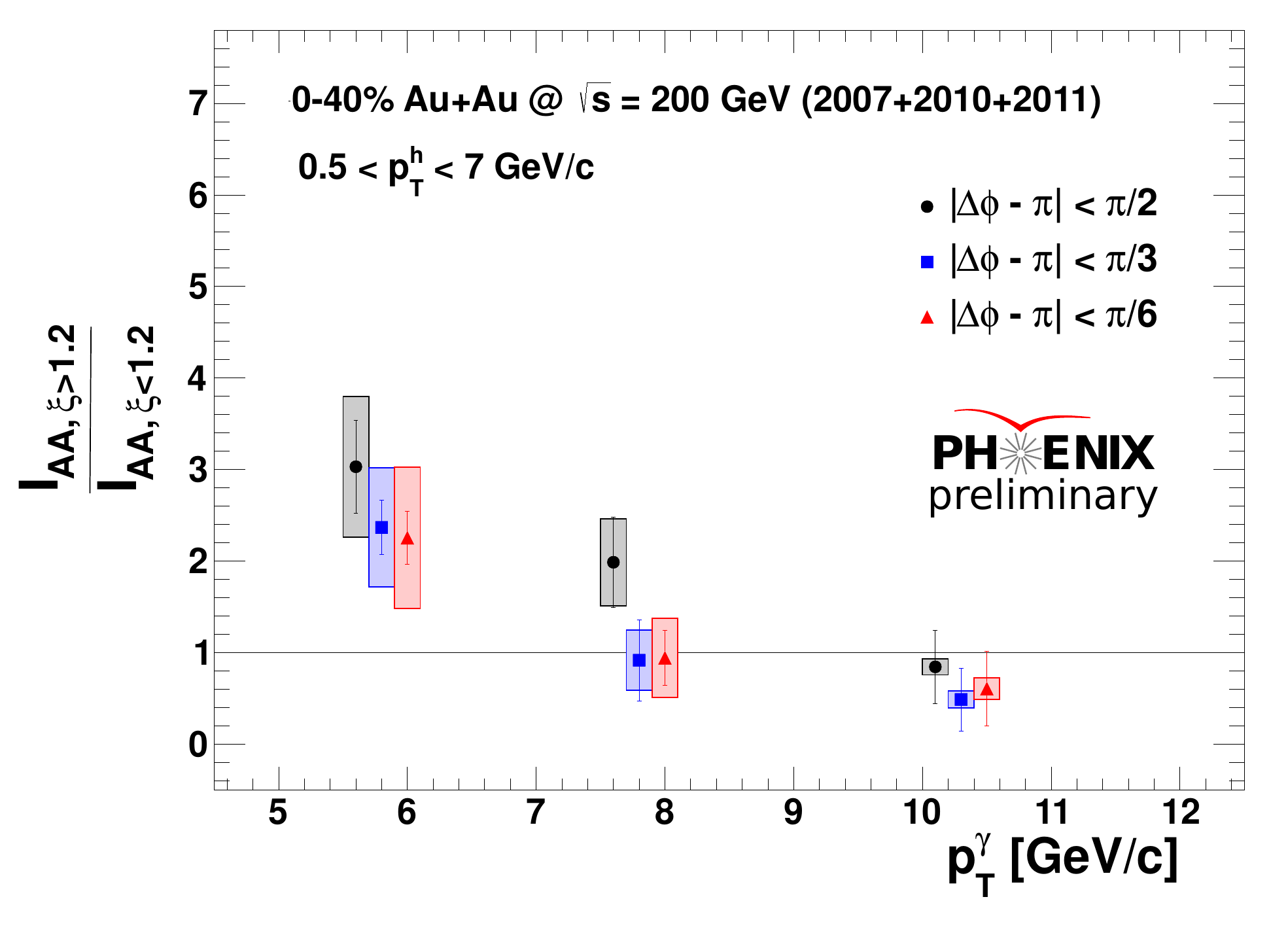}
\caption {Ratios of $I_{AA}$ as a function of trigger $p_T$, for three different away-side integration ranges.}
\end{figure}

\section{Summary}
In conclusion, PHENIX has measured direct photon-hadron correlations in both Au+Au and d+Au collisions at $\sqrt{s}$ = 200 GeV. Comparisons to model calculations suggest that the $I_{AA}$ transition does not occur at a fixed $\xi$ value. Models that include medium response can describe the data reasonably well. $I_{dA}$ is in agreement with unity given the current uncertainties of the measurement, suggesting no CNM effect is has been observed.\newline



%% References

\nocite{*}
\bibliographystyle{elsarticle-num}
\bibliography{jos}



\end{document}